# A Universal Hurricane Frequency Function

Robert Ehrlich

George Mason University, Fairfax, VA

December 28, 2009

______________________

*Corresponding author address: Robert Ehrlich, Physics & Astronomy Department, George Mason University, Fairfax, VA 22030.*

E-mail: rehrlich@gmu.edu

ABSTRACT

Evidence is provided that the global distribution of tropical hurricanes is principally determined by a universal function H of a single variable z that in turn is expressible in terms of the local sea surface temperature and latitude. The data-driven model presented here carries stark implications for the large increased numbers of hurricanes which it predicts for a warmer world. Moreover, the rise in recent decades in the numbers of hurricanes in the Atlantic, but not the Pacific basin, is shown to have a simple explanation in terms of the specific form of $H(z)$, which yields larger percentage increases when a fixed increase in sea surface temperature occurs at higher latitudes and lower temperatures.

# 1. Introduction

There are numerous factors that either promote or inhibit the formation of hurricanes or tropical storms, and it is well-known that two of them have special significance: the sea surface temperature SST (or simply T) and the latitude $\phi$ at the time and place of storm formation. The importance of latitude is that it governs the strength of the Coriolis Effect, a principal factor in creating some initial vorticity, while the SST is the source of an updraft of air that can create a tropical low. Here, we assert more specifically that the probability density of a hurricane or tropical storm forming can be expressed in terms of a simple mathematical function of those two variables, $H(T,\phi)$. Alternatively, we can even write H in terms of a single combined variable $z = (T - T_0)\sin^{1/2}|\phi|$, where T is the SST, $T_0$ is a threshold value of 25.5$^0$C, and $\phi$ is the latitude. The functional form of H(z) most consistent with the data is found to be a simple power law, H(z) = Cz$^n$, where C = 0.00073, and n = 3.5$\pm$0.5, for most regions of the globe. This result appears to be independent of time, and location, with the exception of those regional departures.

The study reported here rests on four key assumptions.

*a. Existence*

There exists a probability function H which describes whether a hurricane (also referred to as a cyclone or typhoon) forms. Furthermore, we assume that H depends on numerous variables, of which only the SST and the latitude play a universal role, with the other ``secondary'' variables, being restricted in time or space.

*b. Secondary variables*

The secondary variables influencing H do so in a way that simply multiplies it by a constant factor that is regionally and temporally limited, and can be explained in terms of previously studied regional or oscillatory phenomena that enhance the value of H by a specific factor over that spatial or temporal region.

*c. Data-driven*

The function H can be found using the data on recorded hurricanes, without requiring a fundamental understanding of the basic physics describing exactly how a hurricane forms. This latter subject is one of current research interest, and the clue offered by the specific functional form of H presented here may advance that search for understanding. Since the function H is justified by appealing to data rather than fundamental theory, the result is a model rather than a theory.

*d. Validation*

Given a specific form of H derived from the data, we can test the model by cross-checking it for consistency using the data, and more importantly, by seeing whether future hurricane data agrees with the model.

## 2. Methods

This work has relied on the data sets maintained by NOAA for the Atlantic and Eastern Pacific regions, and by the Joint Typhoon Warning Center (JTWC) Task Force of the U.S. Navy and Air Force for the World's other oceans. These data sets show both the reconstructed SST by month in $2^0$ x $2^0$ latitude-longitude bins for each month from 1854 to the present, and the tracks of tropical storms and hurricanes occurring over the period 1945 to the present for storms in the Western (Asian) Pacific and Indian Oceans, including the Southern Hemisphere, as well as those of Atlantic and Eastern Pacific. For the present study we focus only on the latitude-longitude location for the first point on the track of each storm (assumed to be its point of origin), and we have included both tropical storms as well as hurricanes, which we henceforth do not distinguish.
The era of meteorological satellites began in the early 1960's, so it might be expected that geographic distributions of points of origin of tropical storms would be less biased for storms recorded since 1960 -- a suggestion which is confirmed by the observed depopulation of storms in regions far from shipping lanes of Atlantic storms before 1960 compared to later ones. Thus, in order to use a data set that is not clearly biased in terms of geographic distribution, we only use storms since 1960 in finding H(z).

Tropical storms can be thought of as heat engines which derive their energy from the warm surface temperature of the ocean -- or more precisely the temperature difference between the warm ocean and the much lower temperature at high altitude in the atmosphere, which can represent a differential of around $100^0C$. It has been observed that tropical storms do not arise unless the SST, T, exceeds some threshold value $T_0$ cited in various sources as being 26 or $27^0C$. The other key variable in storm formation besides SST is the latitude, which is important because the horizontal component of the Coriolis force varies as its sine. The Coriolis force produces some initial vorticity that helps the storm self-organize. Thus, tropical storms do not form very close to the equator, and they have opposite senses of rotation in the two hemispheres.

In searching for a function H, we make use of the above well-known aspects of tropical storm formation, and hence assume that the relevant variables entering the distribution H involve the quantities $T - T_0$ and $\sin \phi$. Since it is much easier searching an unknown function of a single variable than one involving two variables, we divide the search process into two steps: first, finding a single combined variable z expressed in terms of $T - T_0$ and $\sin \phi$ and second finding an appropriate functional form of z that best fits the data. One plausible combination of the variables when $T - T_0 > 0$ is $z = (T - T_0)^a \sin^b |\phi|$, which vanishes when either $T = T_0$ and $\phi = 0$, although one might well imagine other possible choices. The second part of the search for the form of the function can be done visually by trial and error (for any given trial values of a and b) because we want the resulting H(z) generated using the data themselves to satisfy these six requirements:

*a. Threshold*

H (z) must have the correct threshold behavior. This requirement is guaranteed by our choice of z for all a and b, as long as H(z) vanishes when z approaches zero.

*b. Symmetry*

H (z) must be a symmetrical function between the hemispheres even though the actual numbers of storms is not symmetric.

*c. Monotonicity*

H (z) must be a monotonic function of z since we expect H to increase continually as either $T = T_0$ or $\phi$ increase – although we cannot rule out a priori the possibility that it levels off at some saturation value for large z.

*d. Uniformity*

H (z) must have the same functional form in each basin across the globe.

*e. Normalization*

H(z) ideally should also have the same normalization in each basin, or if not, the variations should be explainable in terms of known regional phenomena.

*f. Zeroes*

H(z) must not predict storms occurring in any regions of the globe where none are observed, e.g., in the South Atlantic.

Let us now explain exactly how H(z) is deduced from the data themselves. The point of origin of each recorded storm in the data base places it in a specific latitude-longitude cell (taken here to be 4 x 4 degrees), and since the time of occurrence of that storm is known, we can use the SST-record to find the T in that cell when the storm formed. Given a particular trial definition relating z to T and $\phi$, we can then compute the z-value associated with each storm. The z-values are binned, and we simply count the number of storms $N_j$ in the jth bin between $z_j$ and $z_{j+1}$. We then use the SST-record to obtain T for a given month for each latitude-longitude cell to find the number of months during the 47-year record from 1960-2007 that a given cell had a temperature T. Finally combining that T with the $\phi$ values of each cell, we find the number of months $M_j$ that cells in a given geographic basin (such as the Atlantic) had a specific binned z-value, $z_j$. The computed value of H for that $z_j$ simply equals $N_j/M_j$, which represents the observed number of storms per month in 4x4 degree latitude-longitude cells for a given z. Since these numbers are always less than one, it is more appropriate to think of them as the probability of having a storm per month. Once we have deduced from the data a set of H-values with associated uncertainties (discussed later), we can proceed to find the functional form of H based on a best fit to the data-derived values. Given the fairly stringent requirements on the form of H(z) discussed earlier, the search for an H(z) that fits the data is fairly easy, given any trial choice of a and b used to define the variable z.

For example, we have found that the simplest choice: $z = (T - T_0)\sin|\phi|$ is found to give poor results, while the choice: $z = (T - T_0)\sin^{1/2}|\phi|$, gives good ones as we shall see in the next section. Based on chi square for the Asian Pacific basin the best fitting exponent b in $z = (T - T_0)^a \sin^b |\phi|$ is actually b=0.49 ±0.01.) One interesting property of our choice of $z = (T - T_0)^a \sin^b |\phi|$ is that it yields a common temperature threshold $T_0$ for all latitudes, which may seem surprising. Other possible choices such as $z = (T - T_0)^a - c\sin^b |\phi|$ would not have this feature, but they would introduce one more free parameter, and would imply that H(z) is no longer separable into functions of T and $\phi$. Were our choice not found to yield a good fit to the data, one would need to explore such more complex definitions of z.

## 3. Results

In order to test how well the data from various regions of the world agree with a power law in z, i.e., $H(z) = Cz^n = (T - T_0)^n \sin^{n/2} |\phi|$, we have relied heavily on the data for the Asian Pacific basin. These data constrain H(z) far more than Atlantic data, since (a) the Asian Pacific basin includes over five times as many storms as the Atlantic basin, (b) it includes storms over a larger range of latitudes (31 bins in z versus only 17 for Atlantic data), and (c) the Asian Pacific includes appreciable numbers of storms in both hemispheres, whereas fewer than 1% of Atlantic storms are in the Southern Hemisphere. We show in fig. 1 H(z) values derived from the data for the Asian Pacific basin, where we artificially display Southern Hemisphere data points as having negative z-values, so as to demonstrate the degree of symmetry for the data-derived function H(z). The data may be seen to agree well with a n=3.5 $\pm$ 0.5 power law for H(z), with C= 0.00073$\pm$ .00006, and $T_0$ = 25.5 $^0$C. The chi square for the fit is 29.2 for 30 d.o.f. (P = 51 %). Before discussing the quality of fit for data in the other hurricane basins, we digress to consider the size of the error bars used in fig.1, which directly affects the quality of the fits.

*a. Size of Uncertainties*

Although one often only displays error bars on the dependent variable only, here the independent variable z also has a significant uncertainty that must be taken into account. The former uncertainty arises because of the statistical errors inherent in the computation of H for each data point. Thus, to find $\Delta H$ we assume that $N_j$ and $M_j$ are uncertain by $\sqrt{N_j}$, and $\sqrt{M_j}$ respectively, yielding for $\Delta H = \sqrt{(1+H)H/M}$. The uncertainty in the independent variable z is not merely $\pm$ 0.05, as might be expected based on half the bin width in z, but instead the larger value $\pm$ 0.15. The reason for this enlargement of the z-errors is that from the definition of $z = (T - T_0)\sin^{1/2}|\phi|$, we see that because the data is binned in 4 by 4 degree bins and also in increments of 0.25 $^0$C in T, we need to use an uncertainty (due to binning) of $\Delta\phi = 2^0$ and $\Delta T = 0.125^0 C$. When these errors are propagated and added in quadrature to find the uncertainty in z using representative values of T and $\phi$ for storms, we find that $\Delta z \approx \pm 0.15$.

Finally, since the quantity H(z)/z$^n$ should be a constant if H(z) satisfies a z$^n$ power law, we may use standard error propagation methods to find the variance in H(z)/z$^n$ by combining the separate uncertainties in z and H according to:

$$\sigma^2 = \left(\frac{nH}{z^{n+1}}\Delta z\right)^2 + \left(\frac{\Delta H}{z^n}\right)^2$$

Note that it is this resultant uncertainty that is used to compute chi square under the requirement that H(z)/z$^n$ remains constant, and thus: $\chi^2 = \sum \frac{1}{\sigma^2}\left(\frac{H(z)}{z^n} - C\right)^2$

*b. The Other Hurricane Basins*

Let us now consider how well the data from other basins fits H(z), starting with the Eastern Pacific, principally the region off the Southern coast of Mexico -- a region of very frequent storms.  This data (the filled triangles in fig. 2) appears to be at odds with what was seen in the Western (Asian) Pacific basin in two important respects.  First, although most of the data again lies on a z$^{3.5}$ power law (upper) curve, its normalization is 3.6 times higher than for the Western Pacific data. Second, three data points with the largest z-values lie on the lower curve having the original normalization shown for fig. 1.

As it happens the Easter Pacific is one area significantly affected by the Madden-Julian Oscillation (MJO). The MJO is a large-scale quasi-periodic modulation of tropical winds that travels eastward from Asia to America.  It has been previously shown that hurricane activity in the Eastern Pacific is around four times more likely during the phase of the oscillation associated with ascending rather than descending wind flows – see Maloney (2000). Furthermore, the MJO oscillation leads to a significant modulation of the SST by around 0.5 $^0$C during the cycle, with the phase of the increased hurricane activity being associated with a reduction in SST – see Maloney (2001} These preceding findings can help explain the otherwise anomalous results of fig.2. Given that most storms in the Eastern Pacific occur in a fairly narrow range of latitudes off the Southern coast of Mexico, larger z values are equivalent to higher temperatures. We thus see that the three filled triangle data points for the highest z-values are consistent with a $H(z) = Cz^{3.5}$, with the original normalization observed for the Pacific data, while the data corresponding to lower z-values are consistent with $H(z) = 3.6Cz^{3.5}$, given the effect of the MJO in increasing hurricane formation approximately four-fold during the phase associated with lower SST.

A similar regional effect must be invoked to explain the data in the Atlantic basin, for which a narrow band of increased hurricane genesis can be observed between South America and Africa at a latitude range of between 10 and 15 degrees North latitude. As can be seen (open squares in fig. 2), the data in this band again lie on a n=3.5 power law curve with the same factor of 3.6 enhancement that was found for the Eastern Pacific data. As it happens, this narrow band of the Atlantic is during the summer a major portion of the Intertropical Convergence Zone (ITCZ), where winds from the two hemispheres converge, and it creates an area of low atmospheric pressure and the rapid upward convection of moist air -- conditions which are conducive to the formation of hurricanes. The remainder of the Atlantic basin away from this band is also consistent with $H(z) = Cz^{3.5}$ (chi square = 17.1 for 17 d.o.f), with the original normalization -- i.e., no factor of 3.6 multiplier. Thus, the data from all portions of the globe having significant numbers of hurricanes are consistent with this same functional form, and even the same normalization, provided one includes a factor of 3.6 enhancement in those regions subject to constant or quasi periodic influences that have been found previously to increase hurricane formation. Furthermore, the function $H(z)$ does not predict any storms should occur in regions of the globe where none are found.

*c. Temporal Trends*

The results presented until now all deal with the spatial distribution of hurricanes, and the function $H(z)$ was deduced with no direct reference to temporal variations -- except through the correlation between local SST and time. The model can, however, be used to predict temporal variations based on the SST of each latitude-longitude bin at any given time. The procedure is to sum the fitted $H(z)$ over all cells with a given z for that particular time. Therefore, the model (if correct) should agree with the observed temporal variations in numbers of storms over some extended time period. In what follows, we make these comparisons with the data three different ways. One comparison (figs. 3, and 4), shows the numbers of hurricanes predicted to occur each month of the year (averaged over the period 1960-2007) by the model with those actually observed in three particular basins: the North Asian Pacific, the Southern Hemisphere, and the combined Atlantic and Eastern Pacific basins. In most cases the agreement is fairly good, with the Southern Hemisphere probably being the worst. It should be noted that the horizontal error bars extend 50% beyond the half-month associated with the bin width. The reason for increasing these error bars is that it has been found that the process of hurricane genesis from some initial disturbance has a tendency to cluster in time sometimes over a period of several weeks. Gray (1979). Since we want to associate SST with the very start of the process, not the observed first point on a hurricane track, we therefore add an extra $\pm$ 1 week uncertainty to the horizontal error bars.

A second temporal variation we may check is the year-to-year variation in numbers of hurricanes in different basins. The poorest of the comparisons is that for the Asian Pacific basin where the model shows a rise towards the end of the 47-year interval 1960-2007, while the data (fig. 5) shows almost no rise. The extent of the discrepancy can be judged by a comparison of a quadratic fit to the model (dotted curve) that is superimposed on the data. This discrepancy is bothersome, but it has not yet reached a statistically significant level -- at least if one includes $\sqrt{N}$ statistical uncertainties in both the data and the model. The chi square for the fit is a perfectly acceptable 38.0 for 47 d.o.f.) It must also be noted that the Asian basin data is much less reliable than the Atlantic data in terms of time trends, especially for data prior to 1985, according Charles Sampson who has made corrections to this data.-- Sampson (2008}.

In contrast to the Pacific time trend, that for the Atlantic basin agrees very well with the model. Even though the Atlantic data pre-1960 may have missed some storms, some researchers have argued that the number missed is not very significant – see: Mann (2007) Therefore we show the time trends in the data and model all the way back to the beginning of the Atlantic record in 1854. The agreement between data and model over this extended period is fairly good. For example, if one just uses the statistical errors in the data themselves, the chi square is 222 (153 d.o.f.) with a probability of only 0.02%, but when one also includes statistical errors in the model it drops to 105, which is ``too good." (i.e., P = 99.9 %) Furthermore, separate quadratic fits to the data and model for the period 1960-2007 are nearly indistinguishable -- see dotted curve in fig. 6 for the fit to the model.

A final method of comparing data and model time trends is to divide the time interval 1960-2007 into two halves, and ask what the model predicts for the increase from the first half of the interval to the second in each 4x4 degree latitude-longitude cell. One then can look at the observed increase in hurricane activity in each cell over that same time period, and group together all those cells corresponding to given levels of predicted increases (in increments of 10 %), and then see what their average observed increase is. A direct comparison can then be made between the observed and predicted increases -- figure 7. The horizontal error bars correspond to half the bin width in fig 7, and the vertical error bars correspond to statistical variations in the numbers of hurricanes in each case -- which are quite large for very large predicted increases which are present only for a very small number of cells, where very few storms are predicted or observed. If the predicted and observed percentage increases were identical, the data points should lie on a line through the origin with unit slope. Despite the large error bars, it is reassuring that the best fit straight line passing through the origin has a slope $1.25 \pm 0.15$ that is only slightly farther from 1.0 than one standard deviation. In fact the chi square for a line with unit slope (shown dotted) is 15.5, giving an acceptable probability of 26%.

## 4. What constitutes a test of the model?

As previously noted, all three types of time trend comparisons discussed above did not factor into the original selection of H(z) which relied on fitting data that was averaged over time, temperature and latitude for any given z-value, and hence there was no a priori guarantee that good fits to the temporal trends could be assured, or even that the data would be describable using a single variable z. Nevertheless, the good temporal trend fits cannot really be said to validate the model, because for any given \phi, the time and temperature are strongly correlated variables. On the other hand, even if the model is viewed as merely a data-fitting exercise, the finding of a function of a single variable that fits the data reasonably well could have great utility, should the model continue to describe the distribution and numbers of hurricanes in the coming years, as global temperatures continue to rise. The true test of the model therefore will come if the numbers of hurricanes in various basins increase in the manner predicted by the model. In particular, one should be able to see the upturn in the Asian Pacific region seen in the model but not yet in the data. Failure to see any such increase in perhaps three to five years of further rising SST's would be fatal for the model.

It has been shown by others that the trend in Atlantic hurricanes correlates well with mean SST's going as far back as 1850 -- see Mann (2007), but the new element here is that of a universal function of local SST (at the time of hurricane formation) that fits the data world-wide, with regional adjustments in normalization. For example, our model offers a simple explanation why with warmer SST temperatures in recent decades, we have witnessed a significant rise in the number of Atlantic hurricanes, but very little if any rise in Pacific hurricanes -- an inconsistency which has led some researchers to a (probably false) conclusion that SST is not that important in determining how often hurricanes form. The model explains the apparent inconsistency, because in comparison to the Western Pacific, much of the region showing the highest percentage increases in storms in the Atlantic basin occurs in cells at higher latitudes, where SST's are cooler. For a given increment in temperature $\Delta T$, the function $H(z)=Cz^{3.5}$ produces a much larger percentage increase for lower temperatures (above $T_0$ than for higher ones that are further from the 25.5 $^0$C threshold.

## 5. Implications

This is not a paper about global warming, but the implications for a warmer world are stark. While it has previously been shown that destructive potential of hurricanes is likely to increase in a warmer world – see: Emanuel (2005), earlier model predictions on the numbers of hurricanes have been highly inconsistent, with many models even suggesting their numbers might decrease, even as they become more destructive – see:.

Oouchi (2006) and Yoshimura (2006). If our results stand the test of time, they imply not only that the numbers of hurricanes will increase as SST rises, but they will do so in proportion to the 3.5 power of the temperature excess above 25.5 $^0$C for any given location. Thus, consider for example a location where the temperature in a given month is 27.5 $^0$C, or two degrees above the 25.5 $^0$C threshold. An average 2 $^0$C rise in the temperature from global warming during that month would increase the numbers of hurricanes there eleven-fold for an n=3.5 power law.

Hurricanes, are among the worst natural disasters in terms of both lives lost and property damaged, as a result of coastal flooding due to storm surges. An eleven-fold increase in hurricanes at a particular location would just be one part of the story, which would include (1) a potentially larger increase in the total number of hurricanes given the increase in the size of the basin as temperatures rise, (2) an increase in the destructive potential of each hurricane, and (3) an increase in the height of the storm surge due to rising sea levels that would invariably occur in a warmer world.

## References

Madden, R.A., & P. R. Julian, P.R., 1971: Detection of a 40-50 day Oscillation in the Zonal Wind in the Tropical Pacific, J. Atmos. Sci., **28,** 702-708.

Orear, J., 1982: Least Squares when both Variables Have Uncertainties, Am. J. Phys., **50**, 912-916

Maloney, E.D., & D. Hartmann, D., 2000: Modulation of Hurricane Activity in the Gulf of Mexico by the Madden-Julian Oscillation, Science, **287**, 2002-2004

Maloney, E.D., & J. T. Kiehl, 2001: MJO-Related SST Variations over the Tropical Eastern Pacific During the Northern Hemisphere Summer, J. of Clim., **15**, 675-689.

Gray, W. M., 1979: Hurricanes: Their formation, structure, and likely role in the tropical Circulation, Meteorology over the Tropical Oceans, D. B. Shaw, Ed., Roy. Meteor. Soc., 155-218.

Sampson, 2008: Informal correspondence with Charles Sampson of the Joint Typhoon Warning Center.

Mann, M.E., Emanuel, K.A., Holland, G.J., & Webster, P.J., 2007: Atlantic Tropical Cyclones Revisited, Eos, **88**, 349-350.

Emanuel, K., 2005: Increasing Destructiveness of Tropical Cyclones over the Past 30 Years, Nature, **436**, 686-688

Oouchi, K., J.Yoshimura, H. Yoshimura, R. Mizuta, S. Kusunoki, and A. Noda, 2006: Tropical cyclone climatology in a global-warming climate as simulated in a 20km-mesh global atmospheric model: frequency and wind intensity analysis. J. Meteorol. Soc. Japan, **84**, 259-276.

Yoshimura, J., M. Sugi and A. Noda, 2006: Influence of greenhouse warming on tropical cyclone frequency. J. Meteor. Soc. Japan, **84**, 405-428.

Lean, J., 1997: The Sun's Variable Radiation and its Relevance for Earth," Ann. Rev. Astron. Astrophys., **35**: 33--67.

# List of Figures

FIG. 1. Probability density of hurricanes H(z) (times 1000) computed from storm data in the Asian Pacific basin for the period 1960 – 2007 shown as a function of the variable z defined in terms of the sea surface temperature T and the latitude $\phi$ according to $z = (T - T_0)\sin^{1/2}|\phi|$ with $T_0$=25.5 $^0$C, and with Southern Hemisphere data artificially displayed as having negative z-values. H(z) represents the probability of having a storm per month for a given z-interval. The size of the horizontal and vertical error bars is discussed in the text. The data is consistent with the n=3.5 power law curve.

FIG. 2. Probability density of hurricanes H(z) (times 1000) versus z computed from storm data in the Eastern Pacific basin (filled triangles) and for a latitude band of the Atlantic basin described in the text (open squares) for the period 1960 – 2007 Apart from three high-z points in the Eastern Pacific basin, the data is consistent with the n=3.5 power law curve, but with 3.6 times the normalization used in fig. 1 --- solid curve. For clarity, error bars have been omitted for the Atlantic data points.

FIG. 3 Number of storms predicted per month of the year during the period 1960 -- 2007 versus numbers actually observed for the combined Atlantic and Eastern Pacific basins. The model predictions (continuous curve) have been normalized to the data to match the total area.

FIG. 4  Number of storms predicted per month of the year during the period 1960 -- 2007 versus numbers actually observed for the North Asia basin (filled circles) and the Asian Southern Hemisphere (open triangles). The model predictions (continuous curve) have been normalized to the data to match the total area.

FIG. 5  Number of storms predicted per year during the period 1960 -- 2007 versus numbers actually observed for the North Asia basin (open triangles), and the nearly flat trend line fit to the data.  The model predictions (continuous curve) have been normalized to the data to match the total area.

FIG. 6 Number of storms predicted per year during the period 1854 -- 2007 versus numbers actually observed for the Atlantic (filled diamonds).  The model predictions (grey curve) have been normalized to the data.  A quadratic fit to the model is shown for the period 1960 -- 2007.

FIG. 7 Observed percentage increase in numbers of storms versus the predicted percentage increase during the two halves of the 48 year interval 1960 -- 2007.  Each data point groups together all cells across the globe having a predicted increase falling in decadal intervals 0 to 10%, 10 to 20%, ... The data should be consistent with a line through the origin having unit slope shown dotted, whereas the best fit line has a slightly larger slope.

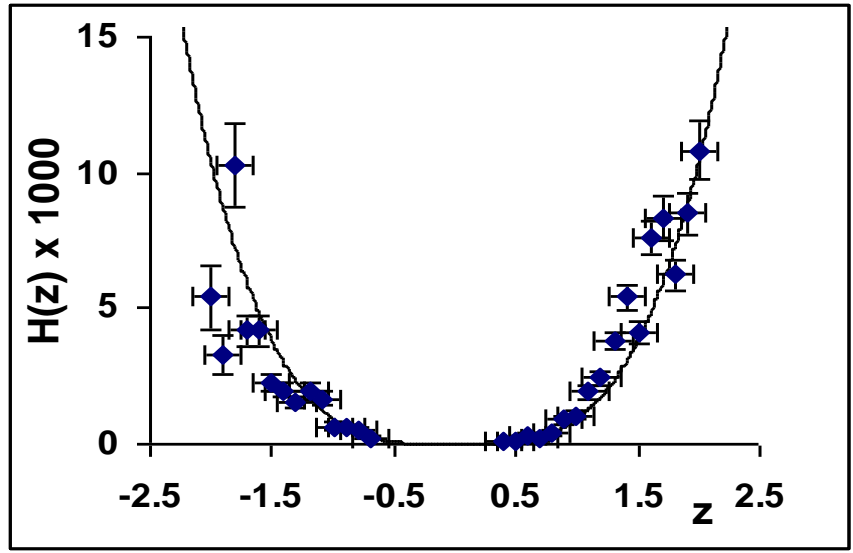

FIG. 1. Probability density of hurricanes H(z) (times 1000) computed from storm data in the Asian Pacific basin for the period 1960 – 2007 shown as a function of the variable z defined in terms of the sea surface temperature T and the latitude $\phi$ according to $z = (T - T_0)\sin^{1/2}|\phi|$ with $T_0$=25.5 $^0$C, and with Southern Hemisphere data artificially displayed as having negative z-values. H(z) represents the probability of having a storm per month for a given z-interval. The size of the horizontal and vertical error bars is discussed in the text. The data is consistent with the n=3.5 power law curve.

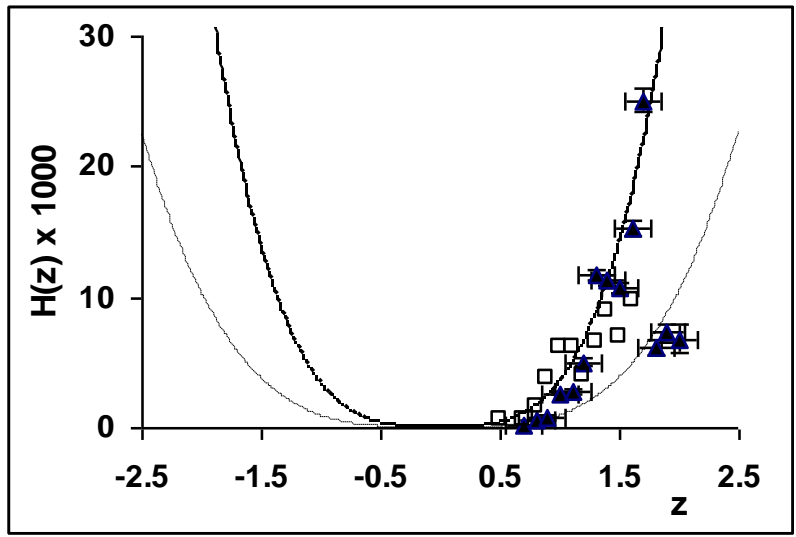

FIG. 2. Probability density of hurricanes H(z) (times 1000) versus z computed from storm data in the Eastern Pacific basin (filled triangles) and for a latitude band of the Atlantic basin described in the text (open squares) for the period 1960 – 2007 Apart from three high-z points in the Eastern Pacific basin, the data is consistent with the n=3.5 power law curve, but with 3.6 times the normalization used in fig. 1 --- solid curve. For clarity, error bars have been omitted for the Atlantic data points.

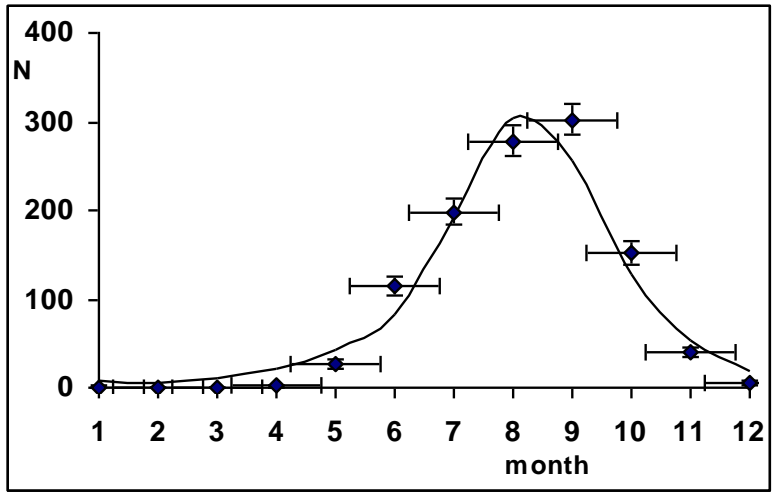

FIG. 3 Number of storms predicted per month of the year during the period 1960 -- 2007 versus numbers actually observed for the combined Atlantic and Eastern Pacific basins. The model predictions (continuous curve) have been smoothed, and have been normalized to the data.}

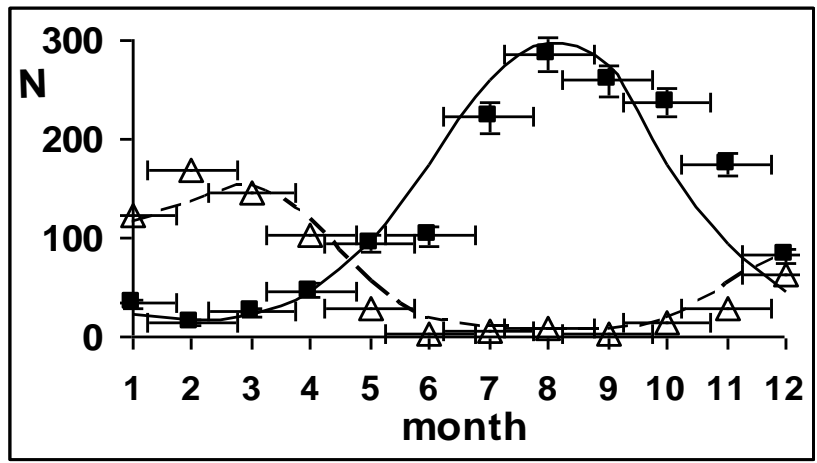

FIG. 4 Number of storms predicted per month of the year during the period 1960 -- 2006 versus numbers actually observed for the North Asia basin (filled circles) and the Asian Southern Hemisphere (open triangles). The model predictions (continuous curve) have been smoothed, and have been normalized to the data.}

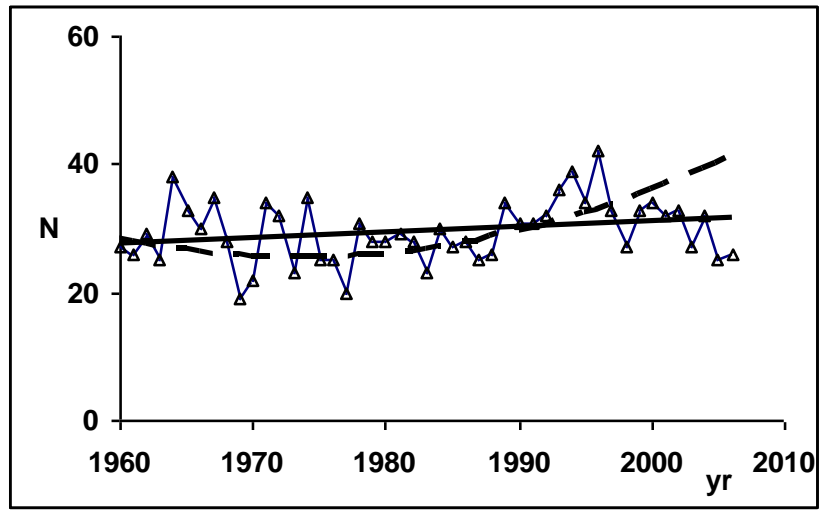

FIG. 5 Number of storms predicted per year during the period 1960 -- 2006 versus numbers actually observed for the North Asia basin (open triangles), and the nearly flat trend line fit to the data. The model predictions (dashed continuous curve) have been smoothed, and have been normalized to the data.

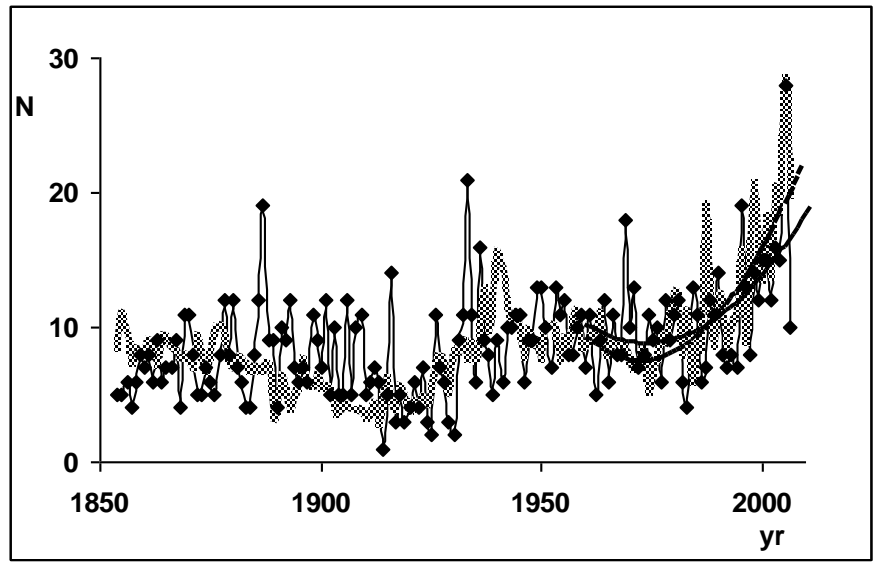

FIG. 6 Number of storms predicted per year during the period 1854 -- 2006 versus numbers actually observed for the Atlantic (filled diamonds). The model predictions (grey curve) have been normalized to the data. A quadratic fit to the model is shown for the period 1960 -- 2006.

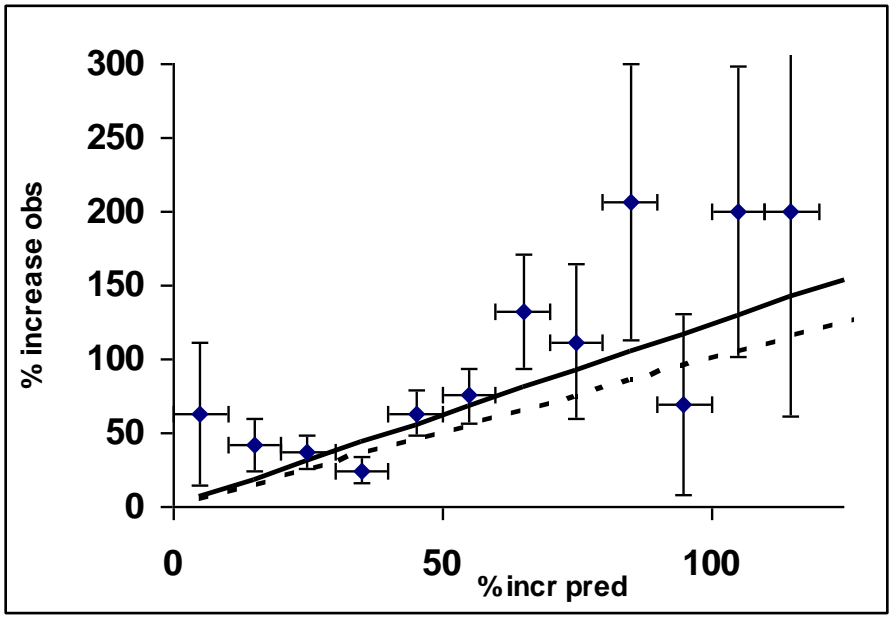

FIG. 7 Observed percentage increase in numbers of storms versus the predicted percentage increase during the two halves of the 47 year interval 1960 -- 2006. Each data point groups together all cells across the globe having a predicted increase falling in decadal intervals 0 to 10%, 10 to 20%, ... The data should be consistent with a line through the origin having unit slope shown dotted, whereas the best fit line has a slightly larger slope.